\documentstyle[epsfig,psfrag,wrapfig]{article}
\newcommand{\be}{\begin{equation}}
\newcommand{\ee}{\end{equation}}
\newcommand{\ba}{\begin{eqnarray}}
\newcommand{\ea}{\end{eqnarray}}

\newcommand{\Mn}{M_{\mbox{\tiny N}}}

\newcommand{\di}{ {\rm d} }

\newcommand{\lsim}{\renewcommand{\arraystretch}{0.4}
                   \begin{array}{c} < \\ \sim \end{array}}
\newcommand{\la}{\langle}
\newcommand{\ra}{\rangle}
\begin{document}
\title{ Sivers vs. Collins effect in azimuthal single spin 
        asymmetries in pion production in SIDIS}
\author{A.~V.~Efremov$^a$, K.~Goeke$^b$, P.~Schweitzer$^c$ \\
        \footnotesize\it $^a$ Joint Institute for Nuclear Research, 
                         Dubna, 141980 Russia\\
        \footnotesize\it $^b$ Institut f\"ur Theoretische Physik II,
                         Ruhr-Universit\"at Bochum, Germany\\
        \footnotesize\it $^c$ Dipartimento di Fisica Nucleare e 
                         Teorica, 
                         Universit\`a degli Studi di Pavia, 
                         Pavia, Italy}
\date{June 2003}
\maketitle
\vspace{-7.5cm}
\begin{flushright}
preprint  RUB-TPII-05/03
\end{flushright}
\vspace{5.5cm}
\begin{abstract}
\noindent
Recently it has been argued that the transverse momentum dependent 
twist-2 Sivers distribution function does not vanish in QCD. 
Therefore both, the Collins and Sivers effects, should be considered in order 
to explain the azimuthal single spin asymmetries $A_{UL}$ in pion production 
in semi-inclusive deeply inelastic lepton scattering of a longitudinally 
polarized target. 
On the basis of presently available phenomenological information on the 
Sivers function we estimate that for those asymmetries $A_{UL}$ in the 
kinematic region of the HERMES experiments the Sivers effect can be 
neglected with respect to the Collins effect. 
It is argued that the same feature holds also
for the COMPASS and CLAS experiments.
This justifies theoretical approaches to understand the 
HERMES data on the basis of the Collins effect only.
\end{abstract}

\paragraph{Introduction.}
Recently, Brodsky, Hwang and Schmidt have shown that leading twist single spin
asymmetries in semi-inclusive deeply inelastic scattering (SIDIS) can arise 
from a rescattering between the struck quark and the target remnant 
\cite{Brodsky:2002cx}. 
Collins has shown \cite{Collins:2002kn} that this rescattering mechanism 
is due to the Sivers effect \cite{Sivers:1989cc}, i.e.\ due to the existence 
of a (naively) T-odd transverse momentum dependent distribution function 
$f_{1T}^\perp(x,{\bf p}_\perp^2)$, correcting his earlier proof that this 
distribution function vanishes \cite{Collins:1992kk} and legitimating 
phenomenological work \cite{Anselmino:1994tv,Boer:1999mm}.
The connection between such rescattering (``final state interactions'') and 
a gauge-invariant definition of $f_{1T}^\perp(x,{\bf p}_\perp^2)$ was further 
elaborated by Belitsky, Ji and Yuan \cite{Belitsky:2002sm}. 

In the light of \cite{Brodsky:2002cx,Collins:2002kn} 
it is not true anymore that the experimental HERMES results 
\cite{Airapetian:1999tv,Airapetian:2001eg,Airapetian:2002mf,Makins}
on azimuthal single spin asymmetries in SIDIS off a longitudinally (with
respect to the beam) polarized target can be interpreted in terms of the 
Collins effect \cite{Collins:1992kk} only. Rather the Sivers effect should 
also be considered. The corresponding tree-level expressions were derived 
by Mulders et al.\  in \cite{Mulders:1996dh,Boer:1997nt}.

In Refs.~\cite{DeSanctis:2000fh,Anselmino:2000mb,Ma:2002ns,Efremov:2000za,Efremov:2001cz,Efremov:2001ia}
it was aimed at a theoretical understanding of the HERMES data
\cite{Airapetian:1999tv,Airapetian:2001eg,Airapetian:2002mf}
in terms of the Collins effect only, relying on the no more valid proof 
\cite{Collins:1992kk} that the Sivers distribution function vanishes.
In principle these works should now be corrected to include the Sivers effect.

In this note we use presently available phenomenological information on 
the Sivers function by Anselmino et al.\ \cite{Anselmino:1994tv} ({\sl cf.} 
\cite{Boglione:1999pz}) to estimate the contribution of the Sivers effect 
to the azimuthal single spin asymmetries in the HERMES longitudinal 
target polarization experiments
\cite{Airapetian:1999tv,Airapetian:2001eg,Airapetian:2002mf}
and find that it can be neglected compared to the Collins effect.
We also argue that in the approach of the present authors 
\cite{Efremov:2001cz,Efremov:2001ia} the neglect of the Sivers effect 
was consistent and justified from a theoretical point of view.

%
        \begin{wrapfigure}{R!}{4.4cm}
        \vspace{-1cm}
        \mbox{\epsfig{figure=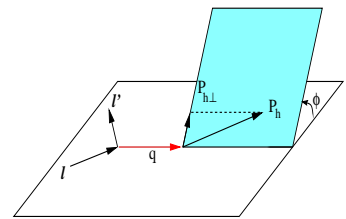,width=4.4cm,height=3cm}}
        \caption{\footnotesize\sl
        Kinematics of the process $lp\rightarrow l'h X$ in the lab frame.}
        \vspace{-0.2cm}
        \end{wrapfigure}
%
%
\paragraph{\boldmath Sivers effect in the HERMES longitudinal 
target polarization experiment.}
Let us consider the process $lp^\pm\rightarrow l'h X$ (see Fig.~1) where 
``$\pm$'' denotes the longitudinal (with respect to the beam) polarization of 
the proton target (``$+$'' means polarization opposite to the beam direction).
With $P$ denoting the momentum of the target proton, 
$l$ ($l'$) denoting the momentum of the incoming (outgoing) lepton and 
$P_h$ the momentum of the produced hadron, the relevant kinematical variables
are $s:=(P+l)^2$, $q= l-l'$ with $Q^2=- q^2$, $x=\frac{Q^2}{2Pq}$, 
$y = \frac{Pq}{Pl}$, $z=\frac{PP_h}{Pq}$.
Let us consider the weighted cross section difference
(cf.\ footnote~\ref{footnote:notation} below)
\be\label{expl-1}
    \Delta\sigma^{\sin\phi \,k_\perp/\la P_{h\perp}\ra}(x) = 
    \int\!\!\di z\,\di y\,\di^2{\bf P}_{h\perp}\,
    	\sin\phi\frac{k_\perp}{\la P_{h\perp}\ra}\left(
    	\frac{1}{S^+}\,
	\frac{\di^5\sigma^+}{\di x\,\di y\,\di z\,\di^2{\bf P}_{h\perp}}
   	-\frac{1}{S^-}\,
	\frac{\di^5\sigma^-}{\di x\,\di y\,\di z\,\di^2{\bf P}_{h\perp}}
    	\right)\;\;,\ee
where $S^\pm$ denotes the modulus of the target polarization and 
$\sigma^{\pm}$ are the corresponding cross sections. The fact that in the 
HERMES experiment the cross sections were weighted without the transverse 
(with respect to the hard photon) momentum $k_\perp=|{\bf P}_{h\perp}|/z$ 
is not relevant for our discussion.

Assuming factorization the cross section asymmetry 
$\Delta\sigma^{\sin\phi \,k_\perp/\la P_{h\perp}\ra}$ was shown in a 
tree-level calculation up to ${\cal O}(1/Q)$ to receive two contributions 
-- one from the longitudinal and one from the transversal
(with respect to the hard photon) component of the target
spin $S$  \cite{Mulders:1996dh,Boer:1997nt}
\be\label{AUL}
        \Delta\sigma^{\sin\phi \,k_\perp/\la P_{h\perp}\ra}(x) 
        =\sigma_{UL}^{\sin\phi \,k_\perp/\la P_{h\perp}\ra}(x)
        +\sigma_{UT}^{\sin\phi \,k_\perp/\la P_{h\perp}\ra}(x)
        \;.\ee
The longitudinal part $\sigma_{UL}^{\sin\phi \,k_\perp/\la P_{h\perp}\ra}$ 
is due to the Collins effect only, while in the transversal part 
$\sigma_{UT}^{\sin\phi \,k_\perp/\la P_{h\perp}\ra}$ both 
Sivers and Collins effect contribute \cite{Boer:1997nt}
\be\label{Def:Col-Siv0}
        \sigma_{UT}^{\sin\phi \,k_\perp/\la P_{h\perp}\ra}(x) = 
        \sigma_{UT}^{\rm Col}(x) +
        \sigma_{UT}^{\rm Siv}(x) \ee
where
\ba\label{clarify-notation}
   	\sigma_{UT}^{\rm Col}(x) \equiv 
	-\;\frac{2}{S}\;\int\!\!\int\!\di y\;\di z\;
	\bigl\la\frac{k_\perp}{\la P_{h\perp}\ra}\sin(\phi+\phi_S)
	\bigr\ra_{OTO,\;{\rm for}\;\phi_S=0} &,& \nonumber\\
   	\sigma_{UT}^{\rm Siv}(x) \equiv
	-\;\frac{2}{S} \;\int\!\!\int\!\di y\;\di z\;
	\bigl\la\frac{k_\perp}{\la P_{h\perp}\ra}\sin(\phi-\phi_S)
	\bigr\ra_{OTO,\;{\rm for}\;\phi_S=0} &,&\ea
in the notation of \cite{Boer:1997nt}, or explicitly
\ba\label{Def:Col-Siv}
        \sigma_{UT}^{\rm Col}(x) &=&
        - 4\pi\alpha^2s \int\di y\; \sin\Theta_S\,2(1-y)Q^{-4}
        \sum_a e_a^2 x h_1^a(x) \la H_1^{\perp(1)a}\ra \;, \nonumber\\
        \sigma_{UT}^{\rm Siv}(x) &=&  
        - 4\pi\alpha^2s\,\frac{\Mn}{\la P_{h\perp}\ra}
        \int\di y\;\sin\Theta_S\,2(1-y+y^2/2)Q^{-4} 
        \sum_a e_a^2 x f_{1T}^{\perp(1)a}(x) \la D_1^a \ra \,, \ea
with\footnote{
	\label{footnote:notation}
	Note, that we use the notation of \cite{Mulders:1996dh,Boer:1997nt}
	with the Collins function normalized with 
	respect to $\la P_{h\perp}\ra$ instead of $m_\pi$, i.e.
	$[H_1^\perp/\la P_{h\perp}\ra]_{\rm here} = $
	$[H_1^\perp/m_\pi]_{\mbox{\tiny\cite{Mulders:1996dh,Boer:1997nt}}}$.
	Correspondingly, it is $\la P_{h\perp}\ra$ which compensates the
	dimension of $k_\perp$ in the weight of the cross section asymmetry
	in Eqs.~(\ref{expl-1},~\ref{AUL},~\ref{Def:Col-Siv0}) and in the
	definition (\ref{kT-moments}). Note also the opposite definition 
	of azimuthal angles in \cite{Boer:1997nt}.}
(recalling the relation ${\bf P}_{h\perp}= - z {\bf k}_\perp$ between the 
fragmenting quark and the produced hadron)
\ba\label{kT-moments}
	H_1^{\perp(1)a}(z) 
	&=& z^2 \int\di^2 {\bf k}_\perp\; 
	    \frac{{\bf k}_\perp^2}{2 \la P_{h\perp}\ra^2}\;
	    H_1^{\perp a}(z, - z {\bf k}_\perp)\;,\nonumber\\
	f_{1T}^{\perp(1)a}(x) 
	&=& \int\di^2 {\bf p}_\perp\; 
	    \frac{{\bf p}_\perp^2}{2\Mn^2}\;
	    f_{1T}^{\perp a}(x,{\bf p}_\perp) \;.
\ea

In Eq.~(\ref{Def:Col-Siv}) $\sin\Theta_S=|{\bf S}_T|/|{\bf S}|$
$=[(4\Mn^2x^2)(1-y-\Mn^2x^2y^2/Q^2)/(Q^2+4\Mn^2x^2)]^{1/2}$ characterizes 
the  transverse (with respect to the photon) component of the target spin
for longitudinal target polarization.  $H_1^\perp$ is normalized according 
to the convention of Refs.~\cite{Mulders:1996dh,Boer:1997nt}. 
In Eq.~(\ref{clarify-notation}) $\phi_S$ denotes the azimuthal angle of the 
target spin around the photon direction with respect to lepton scattering 
plane. For a longitudinal polarization $\phi_S = -\pi$ (for ``$+$'' 
polarization in Eq.~(\ref{expl-1})) such that Sivers and Collins effect
become indistinguishable. When integrating over $y$ and $z$ in 
Eqs.~(\ref{clarify-notation},~\ref{Def:Col-Siv}) one has to consider the cuts 
$W^2>4\,{\rm GeV}^2$ and $Q^2>1\,{\rm GeV}^2$, $0.2<y<0.85$ and $0.2<z<0.7$ 
in the HERMES experiment.

Assuming a Gaussian distribution of transverse momenta\footnote{
        \label{footnote-on-kT}
        This assumption does not contradict the HERMES data, but it 
        is not in agreement with the phenomenological considerations 
        of Collins \cite{Collins:1992kk} or the model calculation of 
	Bacchetta et al.\  in  Ref.~\cite{Bacchetta:2002tk}. 
	However, in a limited $z$-range 
        ($0.2<z<0.7$ in the HERMES experiment) a relation of the kind 
        (\ref{Eq:Assume-Gauss}) can always be assumed to hold with a
        sufficient accuracy for our purposes. 
        Note that strictly speaking in the integration over transverse 
        pion momenta also the experimental cuts should be considered
        ($|{\bf P}_{h\perp}|>50\,{\rm MeV}$ in the HERMES experiments).}
for $H_1^\perp(z,{\bf P}_{h\perp})$ one obtains for the cuts of the HERMES
experiment
\be\label{Eq:Assume-Gauss}
        \la H_1^{\perp(1)}\ra \equiv 
        \int_{0.2}^{0.7}\di z \int\di^2 {\bf P}_{h\perp}\,
        \frac{k_\perp^2}{2\la P_{h\perp}\ra^2}\,H_1^\perp(z,{\bf P}_{h\perp})
        = \frac{\la k_\perp^2\ra}{2\la P_{h\perp}\ra^2} 
	  \int_{0.2}^{0.7}\di z\;H_1^\perp(z)
        = \frac{\la k_\perp^2\ra}{2\la P_{h\perp}\ra^2} 
          \la H_1^\perp\ra \;.
\ee
The $H_1^\perp(z)$ in (\ref{Eq:Assume-Gauss}) is defined by the 
assumption of a Gaussian distribution of transverse momenta. 
It is this quantity which under certain assumptions was extracted 
by the present authors in Ref.~\cite{Efremov:2001cz} from the HERMES 
data \cite{Airapetian:1999tv,Airapetian:2001eg}.
Assuming favoured fragmentation (i.e.\  in the following $\sum_a$ means the 
sum over the corresponding favoured flavours) we obtain for the ratio of 
``Sivers to Collins cross section asymmetry'' the result
\be\label{Ratio}
         \frac{\sigma_{UT}^{\rm Siv}(x)}{\sigma_{UT}^{\rm Col}(x)} =
         \frac{2\Mn\la P_{h\perp}\ra}{\la k_\perp^2\ra}\;
         \frac{\la D_1\ra}{\la H_1^\perp\ra}\;
         \frac{\int\di y\,\sin\Theta_S\,(1-y+y^2/2)/Q^4}
              {\int\di y\;\sin\Theta_S\,(1-y)/Q^4}\;
         \frac{\sum_ae_a^2 xf_{1T}^{\perp(1)a}(x)}{\sum_be_b^2 xh_1^b(x)}\;.
\ee
Considering $\la k_\perp^2\ra=(4/\pi)\la k_\perp\ra^2$ for a Gaussian 
distribution, using $\la k_\perp\ra=\la P_{h\perp}\ra/\la z\ra$ and
inserting $\la P_{h\perp}\ra\approx 0.4\,{\rm GeV}$ and $\la z\ra = 0.41$ 
\cite{Airapetian:1999tv,Airapetian:2001eg}, and the value 
$\la H_1^\perp\ra/\la D_1\ra = (13.8\pm 2.8)\%$ reported in 
\cite{Efremov:2001cz} we obtain for the prefactors in Eq.~(\ref{Ratio})  
in the kinematics of the HERMES experiment
\ba\label{prefactor}
	\frac{2\Mn\la P_{h\perp}\ra}{\la k_\perp^2\ra}\;
        \frac{\la D_1\ra}{\la H_1^\perp\ra} \approx 7.29\;,  && 
	\nonumber\\
	1.03 \lsim \frac{\int\di y\,\sin\Theta_S\,(1-y+y^2/2)/Q^4}
                        {\int\di y\;\sin\Theta_S\,(1-y)/Q^4} 
	     \lsim 1.5 \;.  && \ea
%
        \begin{wrapfigure}{RD}{6cm}
        \vspace{-0.5cm}
        \mbox{\epsfig{figure=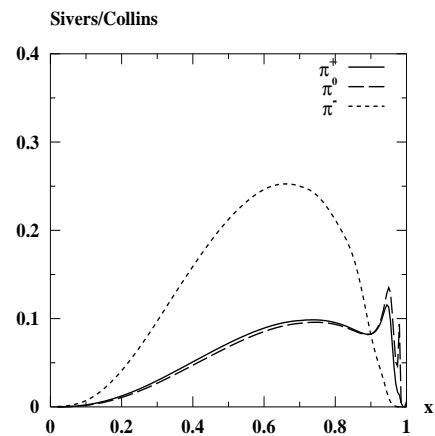,width=6cm,height=6cm}}
        \caption{\footnotesize\sl
          The ``ratio of Sivers to Collins effect'' 
          as defined in Eq.~(\ref{Ratio}) vs. $x$ for the kinematics
          of the HERMES experiment}
        \vspace{-1.2cm}
        \end{wrapfigure}
%

As an estimate for the Sivers function we take the results extracted  
by Anselmino et al.\  in \cite{Anselmino:1994tv} ({\sl cf.} 
\cite{Boglione:1999pz}) from the E704 data \cite{Adams:1991cs} on 
single spin asymmetries in the process $p^\uparrow p\to \pi X$.
If one assumes factorization, there are three possible 
non-perturbative mechanisms which could generate the observed effect:
Collins effect, Sivers effect and a twist-3 mechanism \cite{Efremov:eb}.
Anselmino et al.\  tried to explain the E704 data under the assumption 
that the observed asymmetry is due to the Sivers effect {\sl only}. 
Under this assumption the following fit of the Sivers function to the data 
was reported \cite{Anselmino:1994tv} ({\sl cf.} \cite{Boglione:1999pz})
\ba\label{Sivers-function}
         f_{1T}^{\perp(1)u}(x) &=&  0.81 \, x^{2.70}(1-x)^{4.54}\nonumber\\
         f_{1T}^{\perp(1)d}(x) &=& -0.27 \, x^{2.12}(1-x)^{5.10}\nonumber\\
         f_{1T}^{\perp(1)\bar{q}}(x)&=&0 \; .\ea
The result (\ref{Sivers-function}) refers to a scale which is of the 
order of magnitude of the large transverse momentum of the produced pions 
-- typically $(2-3)\,{\rm Gev}$, i.e.\ comparable to the $Q^2$-region
explored in the HERMES experiment. 
Using the estimate (\ref{Sivers-function}) for the Sivers function 
we obtain for the ratio of ``Sivers to Collins effect'' as defined 
in Eq.~(\ref{Ratio}) the result shown in Fig.~2.
(For the transversity distribution $h_1^a(x)$ we take the prediction of the 
Chiral Quark Soliton Model~\cite{model-h1} LO-evolved to $Q^2=4\,{\rm GeV}^2$ 
as used in \cite{Efremov:2001cz,Efremov:2001ia}.)

As clarified in \cite{Collins:2002kn,Belitsky:2002sm} (see also 
\cite{Boer:2003cm}) the Wilson-link required to ensure gauge invariance of
the expression for the Sivers-function is process-dependent and implies 
opposite signs for $f_{1T}^\perp$ in SIDIS and in the Drell-Yan process. 
The connection between  $f_{1T}^\perp$ in SIDIS and in $p^\uparrow p\to\pi X$
has not been clarified yet.\footnote{
	For a discussion of the processes $p^\uparrow p\to\pi X$ and 
	$e\vec{p}\to \pi X$ at large transverse momenta, which can be 
	described by related (twist-3) functions \cite{Boer:2003cm}, 
	see Ref.~\cite{Koike:2002ti}.}
In Fig.~2 it is assumed that $f_{1T}^\perp$ has the same sign in these 
processes (otherwise the result is to be understood as the modulus of
the ratio of ``Sivers to Collins effect'').

The result in Fig.~2 means that relying on the estimate (\ref{Sivers-function})
the contribution of the Sivers effect to the azimuthal asymmetries from a 
longitudinally polarized target can safely be neglected for the kinematics 
of the HERMES experiment.
We observe that the Sivers effect contributes for $0.023 < x < 0.4$ 
(the HERMES $x$-cuts) about $(2-3)\%$ to the {\sl transversal part} 
of the asymmetry $A_{UL}^{\sin\phi}$ in the case of $\pi^+$ and $\pi^0$.
The transversal part provides a negative and in absolute values smaller 
contribution the total asymmetry $A_{UL}^{\sin\phi}$ compared to the 
{\sl longitudinal part} \cite{Efremov:2001cz}.
(According to our estimates the Sivers effect would contribute about $10\%$ to
the transversal part in the case of $\pi^-$ where, however, unfavoured 
fragmentation effects play a much more important role \cite{Ma:2002ns}. 
In the HERMES experiment $\pi^-$ 
azimuthal asymmetries were found consistent with zero.)
%
It should be noted that the suppression of the Sivers effect with respect to
the Collins effect is not a kinematical effect fore the respective prefactors 
are not small, see Eq.~(\ref{prefactor}).

One could be tempted to interpret (\ref{Sivers-function}) as an upper bound 
for the Sivers function fore it quantifies the right portion of the Sivers 
effect needed to explain the E704 data in terms of this effect only.
However, as already mentioned, there are two more effects namely
the Collins effect and the twist-3 mechanism proposed in \cite{Efremov:eb} 
which could give rise to the effect observed in the E704 experiment. 
One could imagine a situation where the three effects were sizeable,
but contributed to the net result with different signs and partially
canceled each other.
Therefore, we cannot consider the result of Fig.~2 as a strict upper bound
for the contribution of the Sivers effect to the HERMES azimuthal
asymmetries from a longitudinally polarized target. 
Rather we can interpret the result of Fig.~2 as an {\sl indication} that
the Sivers effect is of little importance in the corresponding HERMES 
experiments \cite{Airapetian:1999tv,Airapetian:2001eg,Airapetian:2002mf}.

In this context it is interesting to remark that Anselmino et al.\  also made 
the attempt to understand the E704 data in terms of the Collins effect only, 
observing that in principle this is possible \cite{Anselmino:1999pw}. 
By comparing the $H_1^\perp$ of the present authors ~\cite{Efremov:2001cz} 
   (which explains the HERMES data \cite{Airapetian:1999tv,Airapetian:2001eg}
   by the Collins effect only)
to the  $H_1^\perp$ of  Anselmino et al.\  \cite{Anselmino:1999pw} 
   (which explains the E704 data \cite{Adams:1991cs} 
   by the Collins effect only),
we come to the following conclusion:
The $H_1^\perp$ of the present authors~\cite{Efremov:2001cz} could account for 
roughly half the effect observed in the E704 experiment \cite{Adams:1991cs}. 
In particular, the Collins effect contributes to the E704 data with a 
positive sign\footnote{
	Hereby we assume universality of $H_1^\perp$ in SIDIS and 
	$p^\uparrow p \to \pi X$.  
	In Ref.~\cite{Boer:2003cm} it has recently been argued that there 
	might be no simple relation between the Collins-function in SIDIS 
	and $e^+e^-$ annihilation. (However, see also \cite{Metz:iz}.)
 	The relation between $H_1^\perp$ in 
	SIDIS and $p^\uparrow p\to \pi X$ has not been clarified yet.}.
This would exclude the possibility of a partial cancellation of Sivers and 
Collins effect. Still there is the twist-3 mechanism \cite{Efremov:eb} which 
does not allow us to consider the result in Fig.~2 as a definite bound for 
the contribution of the Sivers effect.

Finally we remark that the attempt to explain HERMES data 
\cite{Airapetian:1999tv,Airapetian:2001eg,Airapetian:2002mf} in terms of the 
Sivers effect {\sl only}, would require a Sivers function one order of
magnitude larger and of opposite sign than in the E704 
experiment (however, cf.\ the discussion above).
So these two experiments are only compatible with each other 
if the Collins effect plays an important role.

Calculations in the quark-diquark models with gluon exchange 
\cite{Brodsky:2002pr,Gamberg:2003ey} suggest a somehow larger 
Sivers-function than the estimate in Eq.~(\ref{Sivers-function}). 
However, it should be noted that even a Sivers-function significantly larger 
(up to an order of magnitude) than (\ref{Sivers-function}) still would yield 
a negligible effect, at least for positive and neutral pions. 
Thus, qualitatively our conclusions are not altered in the light
of the results reported in \cite{Brodsky:2002pr,Gamberg:2003ey}.

\paragraph{CLAS and COMPASS experiments.}
Azimuthal asymmetries will also be studied in the CLAS and COMPASS experiments.
We find that the suppression effect of the Sivers with respect to the Collins
effect in asymmetries from a longitudinally target is only weakly sensitive 
to cuts.
The suppression is stronger with increasing scale because $f_{1T}^{\perp a}$ 
decreases with increasing scale more rapidly than $h_1^a$ 
\cite{Henneman:2001ev}.
This means that dedicated experiments with longitudinally polarized 
targets at CLAS and COMPASS can also be interpreted solely on the basis
of the Collins mechanism. (Predictions for CLAS are presented by the present 
authors in \cite{Efremov:2002ut} and those for COMPASS will be given 
elsewhere).

\paragraph{Sivers effect in SIDIS with transversely polarized target.}
In the case of a longitudinally polarized target both Sivers and Collins 
effect contribute.
In contrast, a transversely polarized target allows a clean separation
of these effects by using appropriate weights \cite{Boer:1997nt}
\ba\label{AUT-Sivers}
        A_{UT}^{\sin(\phi\mp\phi_s)\,k_\perp}(x) &=&
        \frac{
        \int\!\!\di z\,\di y\,\di^2{\bf P}_{h\perp}\sin(\phi\mp\phi_s)\,
	k_\perp\,
    	\left(
	\frac{1}{S^\uparrow}\,
   \frac{\di^5\sigma^\uparrow  }{\di x\,\di y\,\di z\,\di^2{\bf P}_{h\perp}}-
	\frac{1}{S^\downarrow}\,
   \frac{\di^5\sigma^\downarrow}{\di x\,\di y\,\di z\,\di^2{\bf P}_{h\perp}}
	\right)}{
   \frac{1}{2}\int\!\!\di z\,\di y\,\di^2{\bf P}_{h\perp}
	\left(
   \frac{\di^5\sigma^\uparrow  }{\di x\,\di y\,\di z\,\di^2{\bf P}_{h\perp}}+
   \frac{\di^5\sigma^\downarrow}{\di x\,\di y\,\di z\,\di^2{\bf P}_{h\perp}}
	\right)}\nonumber\\
   &\propto&
   	\cases{f_{1T}^\perp D_1 & for ``$-$'' (Sivers effect) \cr
               h_1 H_1^\perp    & for ``$+$''(Collins effect),}\ea
where $k_\perp=|{\bf P}_{h\perp}|/z$.
In the case of transverse polarization the azimuthal angle of the spin 
vector differs from event to event and has to be determined from the data.
More specifically the result for the asymmetry reads \cite{Boer:1997nt}
\ba\label{AUT}
        A_{UT}^{\sin(\phi-\phi_s)\,k_\perp/\Mn}(x) = 
        \frac{2\int\di y\;\cos\Theta_S\,(1-y+y^2/2)Q^{-4} 
             \sum_a e_a^2 x f_{1T}^{\perp(1)a}(x) \la D_1^a\ra}
             {\int\di y\;(1-y+y^2/2)Q^{-4}
             \sum_b e_b^2 x f_1^b(x)\la D_1^b\ra}\ea
Let us estimate the azimuthal asymmetry 
$A^{\sin(\phi-\phi_s)\,k_\perp/\Mn}_{UT}$ on the basis of the results 
(\ref{Sivers-function}) from Ref.~\cite{Anselmino:1994tv}.\footnote{
	Such an estimate has already been given in \cite{Boglione:1999pz} 
	in Fig.~5 and Fig.~6, which show respectively the numerator and 
	denominator of Eq.~(\ref{AUT}) (without $z$-average) as functions 
	of $x$ and $z$ in 3-D plots.} 
Assuming favoured fragmentation we obtain the result shown in Fig.~3a. 
%
\begin{figure}[t!]
\psfrag{x}{\boldmath $x$}
\psfrag{AUT(Sivers) proton}{\boldmath 
        $A_{UT}^{\sin(\phi-\phi_s)\,k_\perp/\Mn}(x)$}
\psfrag{AUTun}{\boldmath 
        $A_{UT}^{\sin(\phi-\phi_s)}(x)$}
\begin{tabular}{ccc}
        \includegraphics[width=5.5cm,height=5cm]{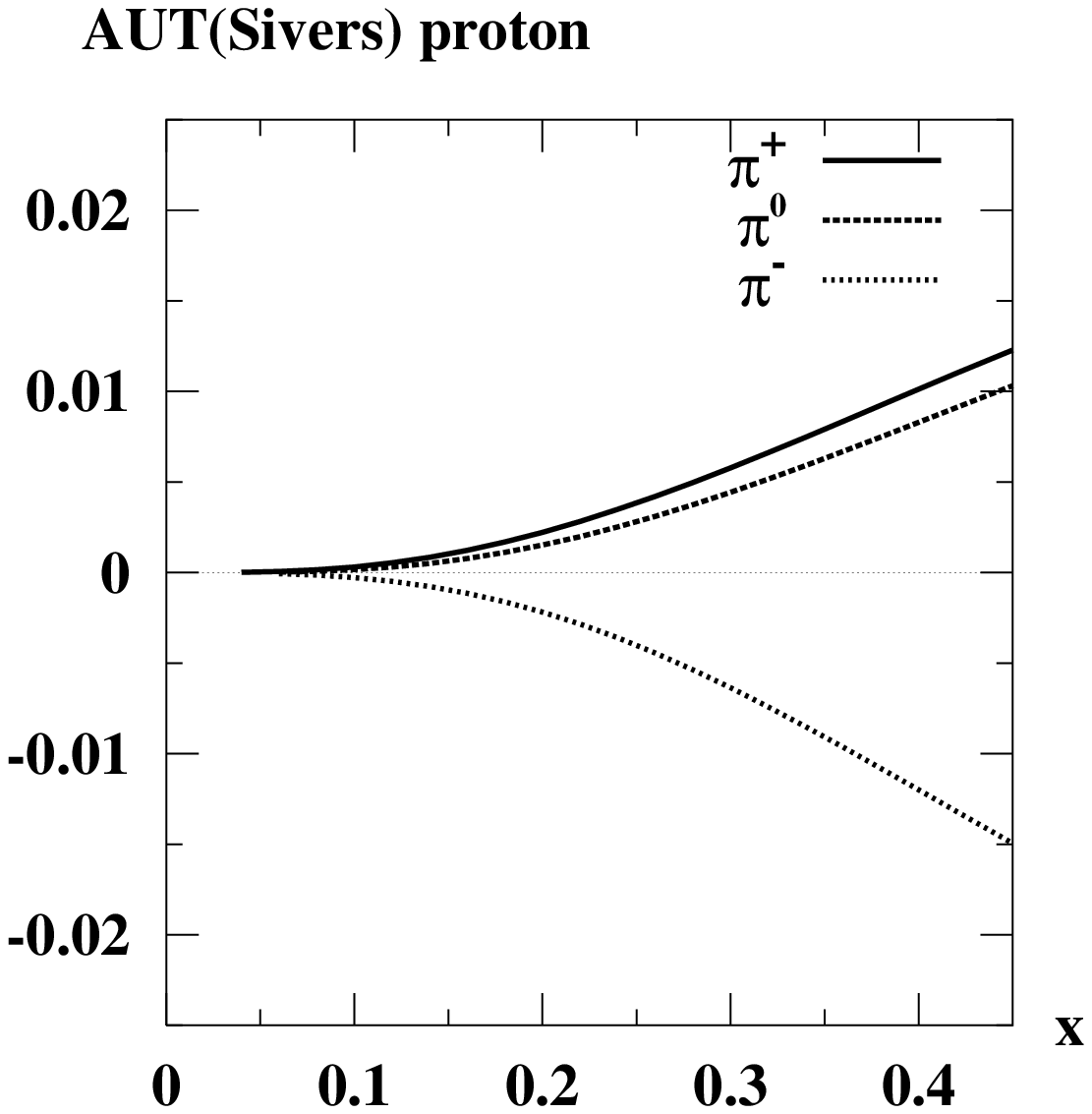} &
	$\phantom{X}$ &
        \includegraphics[width=5.5cm,height=5cm]{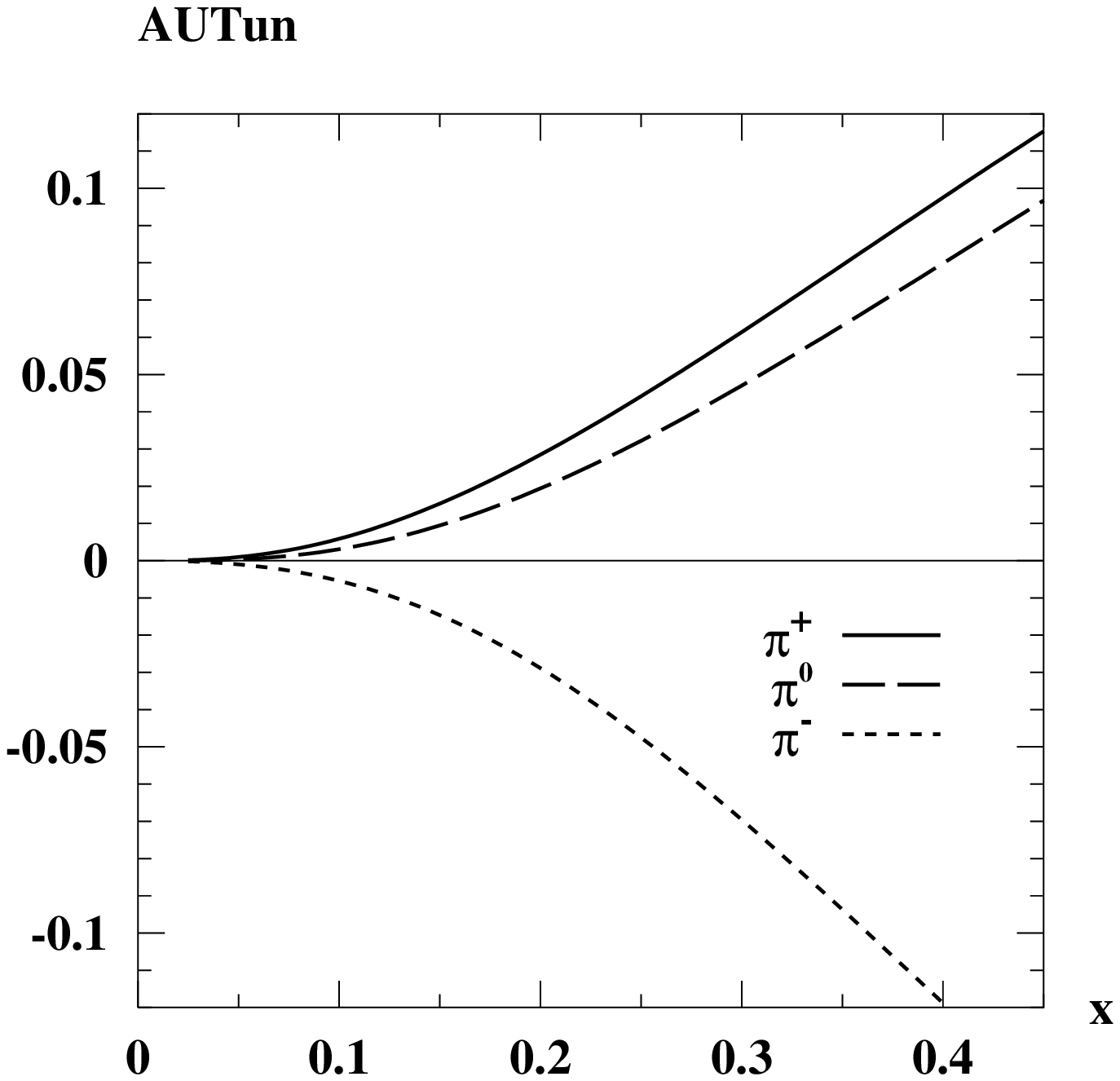}\\
        {\bf a}& &{\bf b}
\end{tabular} 
        \caption{\footnotesize \sl
          (a)
          The single spin azimuthal asymmetry 
          $A_{UT}^{\sin(\phi-\phi_s)k_\perp/\Mn}(x)$ for $\pi^\pm$ and $\pi^0$ 
          for the HERMES kinematics as a function of $x$. Note that the
          weight $k_\perp/\Mn$ provides an artificial suppression, see text.
          \newline
          (b) 
          The single spin azimuthal asymmetry  
          $A_{UT}^{\sin(\phi-\phi_s)}(x)$ (i.e. weighted without the factor 
	  $k_\perp/\Mn$) for the HERMES kinematics as a function of $x$.}
\end{figure}
%
%

The $k_\perp$-weighted asymmetry 
$A^{\sin(\phi-\phi_s)\,k_\perp/\Mn}_{UT}$ is about $1\%$.
However, this does not mean that the effect itself is small
because the $k_\perp/\Mn$-factor in the weight introduces
an artificial suppression of the effect.
In \cite{Anselmino:1994tv,Boglione:1999pz} the result
$\la k_T^2(x)\ra^{1/2} = 0.47 x^{0.68}(1 - x)^{0.48}\Mn$ was 
used from a parton model based analysis \cite{Jackson:1989ph}.
This result implies that $\la k_T^2(x)\ra^{1/2} \lsim 0.2\Mn$ for all $x$.
In order to compare the effect ``more directly'' to the asymmetries 
$A^{\sin\phi}_{UL}\sim (3-4)\%$ measured by HERMES
\cite{Airapetian:1999tv,Airapetian:2001eg,Airapetian:2002mf} we consider 
the asymmetry weighted without the factor $k_\perp/\Mn$, which we estimate 
as follows (cf.\  Appendix)
\be\label{unwAUT}
        A_{UT}^{\sin(\phi-\phi_s)}(x) \approx
        \frac{2\Mn}{\la k_T \ra}\;A_{UT}^{\sin(\phi-\phi_s)k_T/\Mn}(x)\;.\ee
We roughly approximate $\la k_T \ra \approx \la k_T^2\ra^{1/2}$ and 
take $\la k_T^2\ra= \la k_T^2(x)\ra$ from \cite{Jackson:1989ph}.
(It is consistent and necessary to use $f_{1T}^\perp$ from 
\cite{Anselmino:1994tv,Boglione:1999pz} in connection with $\la k_T^2\ra$ 
from \cite{Jackson:1989ph} because the latter was used explicitly in 
Ref.~\cite{Anselmino:1994tv} to fit $f_{1T}^\perp$ under the 
above-mentioned assumptions.)

The result for the asymmetry weighted without the factor $k_\perp/\Mn$, 
$A_{UT}^{\sin(\phi-\phi_s)}(x)$, is shown in Fig.~3b.
We see that the effect itself is not small.
However, the corresponding (non-$k_\perp$-weighted)
$A_{UT}^{\sin(\phi+\phi_s)}(x)\propto h_1^a H_1^{\perp a}$ Collins effect 
asymmetry is larger, namely of ${\cal O}(20\%)$ \cite{Efremov-in-prep}.

Calculations based on a quark-diquark model approach yield an 
$A_{UT}^{\sin(\phi-\phi_s)}(x)$ of comparable magnitude at large (in the 
HERMES kinematics) $x\sim 0.3$, but substantially more sizeable than the 
result in Fig.~3b in the region $x\lsim 0.2$ \cite{Gamberg:2003ey}.
We emphasize that the results in Figs.~3a and 3b can only 
be viewed as rough estimates with the following significance: 
If HERMES measured $A_{UT}^{\sin(\phi-\phi_s)k_\perp/\Mn}$ or
$A_{UT}^{\sin(\phi-\phi_s)}$ of comparable order of magnitude
as the results in Figs.~3a and 3b, then the arguments and estimates 
given in the context of Fig.~2 would experimentally be justified.

\paragraph{\boldmath Comment on the calculations of 
           $A_{UL}$ by the present authors.}
In Refs.~\cite{Efremov:2001cz,Efremov:2001ia} it was aimed at describing 
the HERMES data \cite{Airapetian:1999tv,Airapetian:2001eg,Airapetian:2002mf}
in terms of the Collins effect only. 

The approach of \cite{Efremov:2001cz,Efremov:2001ia} consists 
in combining results from the instanton model of the QCD vacuum 
\cite{Diakonov:1983hh} and from the chiral quark-soliton model 
\cite{Diakonov:1987ty,Christov:1995vm} for the nucleon chirally 
odd distribution functions $h_1^a(x)$ and $h_L^a(x)$ 
\cite{model-h1,Dressler:1999hc}.
This approach is consistent because the chiral quark-soliton model
was derived from the instanton vacuum model. 
A small parameter which played an important role in this derivation is the 
instanton packing fraction --  the ratio of the average instanton size $\rho$ 
to the average separation $R$ of instantons (in Euclidean space-time).
Numerically $\rho/R \sim 1/3$ with $\rho \approx (600\,{\rm MeV})^{-1}$.

The neglect of Sivers function $f_{1T}^{\perp a}$ 
in the approach of Refs.~\cite{Efremov:2001cz,Efremov:2001ia} is fully 
consistent for the following reason.
In Ref.~\cite{Pobylitsa:2002fr} Pobylitsa showed that in a large class of 
chiral models ``$T$-odd'' distribution functions are strictly zero.
In particular in the chiral quark-soliton model $f_{1T}^{\perp a}=0$.
This can be understood by considering that in QCD 
$f_{1T}^{\perp a}$ does not vanish under time reversal only due 
to the non-trivial 
transversal part of the Wilson line entering the definition of 
$f_{1T}^\perp(x,{\bf p}_\perp^2)$ \cite{Collins:2002kn,Belitsky:2002sm}.
Most chiral models are based on quark and Goldstone-boson degrees of freedom,
and the modeling of the Wilson-line is beyond the scope of such models.
  (By itself this does not mean that the Sivers function is necessarily small 
  or even zero in nature. 
  Indeed, other models which explicitly take into account gluonic degrees
  of freedom have no problem with modeling the Wilson-line. One 
  example are the quark-diquark models with gluon exchange of 
  Refs.~\cite{Brodsky:2002cx,Brodsky:2002pr,Gamberg:2003ey}.)

However, if one considers that in the chiral quark-soliton model 
$f_{1T}^{\perp a}=0$ and that this model in a certain sense corresponds 
to the zeroth order in the parameter $\rho/R$ of the instanton model 
\cite{Diakonov:1996sr}, then one arrives at the conclusion that 
$f_{1T}^{\perp a}$ is suppressed in the parameter $\rho/R$. 
This conclusion is drawn here indirectly and should, of course, 
be reinvestigated in the instanton vacuum model using the
methods developed in \cite{Diakonov:1995qy}.

For the calculations of $A_{UL}$ at HERMES by the present authors 
\cite{Efremov:2001cz,Efremov:2001ia} the
information that $f_{1T}^{\perp a}$ is strictly zero in the chiral 
quark-soliton model (and suppressed in the instanton vacuum model)
is, from the theoretical point of view, fully sufficient to neglect 
Sivers effect. 
As we have seen above, this is supported also by phenomenology.

\paragraph{Conclusions.}
It was shown that on the basis of presently available phenomenological
information on the Sivers function \cite{Anselmino:1994tv} the contribution 
of the Sivers effect to single spin asymmetries from a {\sl longitudinally}
polarized target can be neglected with respect to the Collins effect.
This result means that the HERMES data 
\cite{Airapetian:1999tv,Airapetian:2001eg,Airapetian:2002mf}
indeed provide us with first insights into the chirally odd structure
of the nucleon, and that the theoretical efforts to understand these data in 
terms of the Collins effect only \cite{DeSanctis:2000fh,Anselmino:2000mb,Ma:2002ns,Efremov:2000za,Efremov:2001cz,Efremov:2001ia}
were justified.
The same applies to CLAS and COMPASS kinematics, 
which is an encouraging result for these experiments where
(most of the beam-time) longitudinally polarized targets will be used.

Single spin azimuthal asymmetries from {\sl transversely} polarized targets 
allow an unambiguous distinction of the Collins and Sivers effect.
Such asymmetries are presently studied by the HERMES \cite{Makins}
and COMPASS collaborations.
Relying on the information on the Sivers function from 
Ref.~\cite{Anselmino:1994tv} one can estimate the Sivers effect on single spin 
asymmetries from a {\sl transversely} polarized proton target to be about 
$5\%$.  This must not be considered as an absolute prediction, rather 
as a rough indication for the size of the effect, and will serve as a 
thorough experimental test of the considerations presented in this note.
Our predictions for the Collins effect in single spin asymmetries from a 
{\sl transversely} polarized target will be presented elsewhere.

\paragraph{Acknowledgement.}
We are grateful to P.~V.~Pobylitsa and H.Avakian for fruitful discussions.
A.~E.\ is partially supported by INTAS grant 00/587
and RFBR grant 03-02-16816.
This work has partly been performed under the contract  
HPRN-CT-2000-00130 of the European Commission. The work is partially
supported by BMBF and DFG of Germany and by the COSY-Juelich project.

\paragraph{Appendix.}
Explicit formulae for azimuthal asymmetries weighted with an appropriate 
power of transverse momentum were derived in \cite{Boer:1997nt}. 
In asymmetries weighted without an appropriate power of $k_\perp$ 
the transverse momenta in the unintegrated distribution and 
fragmetation functions, in our case $f_{1T}^{\perp a}(x,{\bf k}_T^2)$ and 
$D_1(z,{\bf K}_{T,D}^2)$, remain convoluted. (For a discussion of the meaning 
of unintegrated distribution functions in QCD, see Ref.~\cite{Collins:2003fm}.)
In order to deal with this case 
two approaches have been followed in the literature. 

One approach consists in approximating 
(in the lucid notation of \cite{Boer:1997nt})
\be\label{Eq:app1}
	\bigl\la\sin(\phi-\phi_S)\bigr\ra_{OTO} 
	\approx \frac{2\Mn}{\la k_T\ra}
	\bigl\la\frac{k_\perp}{\Mn}\sin(\phi-\phi_S)\bigr\ra_{OTO}\;.
\ee
This approach was chosen in Refs.~\cite{DeSanctis:2000fh,Ma:2002ns} 
in studies of $A_{UL}$ asymmetries, and in Ref.~\cite{Gamberg:2003ey} 
in studies of the $A_{UT}$ asymmetry which we consider here. 
Eq.~(\ref{unwAUT}) is just the estimate (\ref{Eq:app1}) in a 
different notation.

Another approach consists in directly evaluating the asymmetries with a 
Gaussian ansatz (cf.\  footnote~\ref{footnote-on-kT}) 
\be\label{Eq:app3}
	F(x,{\bf k}_T^2) = F(x)\;
	\frac{\exp(-{\bf k}_T^2/\la{\bf k}_T^2\ra)}{\pi\la{\bf k}_T^2\ra}
\ee
where $F(x,{\bf k}_T^2)$ denotes a generic unintegrated distribution or 
fragmentation function.  The normalization factors in Eq.~(\ref{Eq:app3}) 
are such that $\int\di^2{\bf k}_T F(x,{\bf k}_T^2) = F(x)$ holds.
Under the assumption (\ref{Eq:app3}) we obtain
\ba
	A_{UT}^{\sin(\phi-\phi_s)}(x) 
	&\stackrel{\rm Gauss}{=}& B_{\rm corr} \times 
	\biggl\{\mbox{the result in Eq.~(\ref{unwAUT})}\biggr\}\;,
	\nonumber\\
&&	\label{Eq:app4}
	B_{\rm corr} = \frac{\pi}{4}\cdot \frac{1}{2}\cdot
	\biggl(1+\frac{\la K_{T,D}^2\ra}{\la k_T^2\ra}
	\biggr)^{-1/2}\;.\ea
The transverse momenta of the fragmenting quarks are related to those of the
produced hadrons by $\la K_{T,D}^2\ra \approx \la P_{h\perp}^2\ra/\la z^2\ra$.
Eq.~(\ref{Eq:app4}) can be derived following Ref.~\cite{Mulders:1996dh},
where explicit examples of similar calculations based on the ansatz 
(\ref{Eq:app3}) can be found. 
This approach was chosen in the case of $A_{UL}$ asymmetries in 
\cite{Efremov:2000za,Efremov:2001cz,Efremov:2001ia}.

The appearance of a ``correction factor'' between the heuristic estimate
in Eq.~(\ref{Eq:app1}) and the consistent calculation under a Gaussian
assumption is not surprizing. In the present case one finds an $x$-dependent
$B_{\rm corr} \lsim 0.1$ (for $\la k_T^2\ra$ from \cite{Jackson:1989ph} 
and  $\la K_{T,D}^2\ra \approx \la P_{h\perp}^2\ra/\la z^2\ra$ from the
HERMES experiment). I.e. the second method would yield a substantially
smaller result. 
Both approaches are, of course, heuristic and it is not clear which 
could be more realistic and reliable. In this work we preferred the 
approach based on Eq.~(\ref{Eq:app1}) in order to directly compare to
other calculations reported in literature \cite{Gamberg:2003ey}.

Note that $B_T\neq 1$ means that the estimate (\ref{Eq:app1}) is not 
compatible with a Gaussian distribution of transverse momenta. Therefore
we approximate $\la k_T \ra^2 \approx \la k_T^2\ra$ in the sequence of 
Eq.~(\ref{unwAUT}), as there would be no justification to use, e.g., 
the relation $\la k_T \ra^2 = \pi\la k_T^2\ra/4$ valid for a Gaussian 
distribution only. 

It should be stressed that azimuthal asymmetries offer -- beyond insights 
into the T-odd Collins and Sivers mechanisms -- also the opportunity to learn
about transverse quark momenta in hadrons. An analysis of data from HERMES 
(and other experiments) using both, weights with and without an explicit 
factor $k_T=|{\bf P}_{h\perp}|/z$, could provide valuable phenomenological 
insights. From a strict theoretical point of view, however, the weighting 
with appropriate factors of $k_T$ is preferable \cite{Boer:1997nt}.


\begin{thebibliography}{99}

\bibitem{Brodsky:2002cx}
   S.~J.~Brodsky, D.~S.~Hwang and I.~Schmidt,
   Phys.\ Lett.\ B {\bf 530} (2002) 99 [arXiv:hep-ph/0201296].
\bibitem{Collins:2002kn}
   J.~C.~Collins,
   Phys.\ Lett.\ B {\bf 536} (2002) 43 [arXiv:hep-ph/0204004].
\bibitem{Sivers:1989cc}
   D.~W.~Sivers, 
   Phys.\ Rev.\ D {\bf 41} (1990) 83 [Annals Phys.\  {\bf 198} (1990) 371];
   Phys.\ Rev.\ D {\bf 43} (1991) 261.
\bibitem{Collins:1992kk}
   J.~C.~Collins,
   Nucl.\ Phys.\ B {\bf 396} (1993) 161 [arXiv:hep-ph/9208213].

\bibitem{Anselmino:1994tv}
  M.~Anselmino, M.~Boglione and F.~Murgia,
  Phys.\ Lett.\ B {\bf 362} (1995) 164 [arXiv:hep-ph/9503290].
   M.~Anselmino and F.~Murgia,
   Phys.\ Lett.\ B {\bf 442} (1998) 470 [arXiv:hep-ph/9808426].
\bibitem{Boer:1999mm}
   D.~Boer, Phys.\ Rev.\ D {\bf 60} (1999) 014012 [arXiv:hep-ph/9902255].

\bibitem{Belitsky:2002sm}
   A.~V.~Belitsky, X.~Ji and F.~Yuan,
   Nucl.\ Phys.\ B {\bf 656} (2003) 165
   [arXiv:hep-ph/0208038]. 
   \\
   X.~D.~Ji and F.~Yuan,
   Phys.\ Lett.\ B {\bf 543} (2002) 66 
   [arXiv:hep-ph/0206057]. 

\bibitem{Airapetian:1999tv}
   A.~Airapetian {\it et al.}  [HERMES Collaboration],
   Phys.\ Rev.\ Lett.\  {\bf 84} (2000) 4047 [arXiv:hep-ex/9910062].\\
   H.~Avakian  [HERMES Collaboration],
   Nucl.\ Phys.\ Proc.\ Suppl.\  {\bf 79} (1999) 523.

\bibitem{Airapetian:2001eg}
   A.~Airapetian {\it et al.}  [HERMES Collaboration],
   Phys.\ Rev.\ D {\bf 64} (2001) 097101 [arXiv:hep-ex/0104005].

\bibitem{Airapetian:2002mf}
   A.~Airapetian {\it et al.}  [HERMES Collaboration],
   Phys.\ Lett.\ B {\bf 562} (2003) 182
   [arXiv:hep-ex/0212039]. 

\bibitem{Makins}
   N.~C.~Makins and M.~D\"uren  [HERMES Collaboration],
   Acta Phys.\ Polon.\ B {\bf 33} (2002) 3737.\\
   N.~C.~Makins  [HERMES Collaboration],
   Nucl.\ Phys.\ A {\bf 711} (2002) 41 [arXiv:hep-ex/0209035].

\bibitem{Mulders:1996dh}
  P.~J.~Mulders and R.~D.~Tangerman,
  Nucl.\ Phys.\ {\bf B461} (1996) 197
  [Erratum-ibid.\ {\bf B484} (1996) 538] [arXiv:hep-ph/9510301].
\bibitem{Boer:1997nt}
  D.~Boer and P.~J.~Mulders,
  Phys.\ Rev.\ D {\bf 57} (1998) 5780 [arXiv:hep-ph/9711485].


\bibitem{DeSanctis:2000fh}
  E.~De Sanctis, W.~D.~Nowak and K.~A.~Oganesian,
  Phys.\ Lett.\ B {\bf 483} (2000) 69 [arXiv:hep-ph/0002091];
  V.~A.~Korotkov, W.~D.~Nowak and K.~A.~Oganesian,
  Eur.\ Phys.\ J.\ C {\bf 18} (2001) 639 [arXiv:hep-ph/0002268];
  K.~A.~Oganessian, N.~Bianchi, E.~De Sanctis and W.~D.~Nowak,
  Nucl.\ Phys.\ A {\bf 689} (2001) 784 [arXiv:hep-ph/0010261].

\bibitem{Anselmino:2000mb}
  M.~Anselmino and F.~Murgia,
  Phys.\ Lett.\ B {\bf 483} (2000) 74 [arXiv:hep-ph/0002120].

\bibitem{Ma:2002ns}
  B.~Q.~Ma, I.~Schmidt and J.~J.~Yang,
  Phys.\ Rev.\ D {\bf 66} (2002) 094001 [arXiv:hep-ph/0209114];
  Phys.\ Rev.\ D {\bf 65} (2002) 034010 [arXiv:hep-ph/0110324];
  Phys.\ Rev.\ D {\bf 63} (2001) 037501 [arXiv:hep-ph/0009297].


\bibitem{Efremov:2000za}
  A.~V.~Efremov, K.~Goeke, M.~V.~Polyakov and D.~Urbano,
  Phys.\ Lett.\ B \textbf{478} (2000) 94, hep-ph/0001119.
\bibitem{Efremov:2001cz}
   A.~V.~Efremov, K.~Goeke and P.~Schweitzer,
   Phys.\ Lett.\ B {\bf 522} (2001) 37
   [Erratum-ibid.\ B {\bf 544} (2002) 389] [arXiv:hep-ph/0108213].
\bibitem{Efremov:2001ia}	
   A.~V.~Efremov, K.~Goeke and P.~Schweitzer,
   Eur.\ Phys.\ J.\ C {\bf 24} (2002) 407 [arXiv:hep-ph/0112166];
  Nucl.\ Phys.\ A {\bf 711} (2002) 84;
  Acta Phys.\ Polon.\ B {\bf 33} (2002) 3755 [arXiv:hep-ph/0206267].

\bibitem{Boglione:1999pz}
   M.~Boglione and P.~J.~Mulders,
   Phys.\ Rev.\ D \textbf{60} (1999) 054007, hep-ph/9903354.

\bibitem{Bacchetta:2002tk}
   A.~Bacchetta, R.~Kundu, A.~Metz and P.~J.~Mulders,
   Phys.\ Rev.\ D {\bf 65} (2002) 094021 [arXiv:hep-ph/0201091].

\bibitem{Adams:1991cs}
  D.~L.~Adams {\it et al.}  [FNAL-E704 Collaboration],
  Phys.\ Lett.\ B {\bf 264} (1991) 462.

\bibitem{Efremov:eb}
   A.~V.~Efremov and O.~V.~Teryaev,
   Yad.\ Fiz.\  {\bf 39} (1984) 1517;
   Phys.\ Lett.\ B {\bf 150} (1985) 383.
   \\
   J.~W.~Qiu and G.~Sterman,
   Phys.\ Rev.\ Lett.\  {\bf 67} (1991) 2264;
   Nucl.\ Phys.\ B {\bf 378} (1992) 52.
   \\
   A.~Efremov, V.~Korotkiian and O.~Teryaev,
   Phys.\ Lett.\ B {\bf 348} (1995) 577. 
   \\
   V.~M.~Korotkiian and O.~V.~Teryaev,
   Phys.\ Rev.\ D {\bf 52} (1995) 4775. 
   \\
   J.~W.~Qiu and G.~Sterman,
   Phys.\ Rev.\ D {\bf 59} (1999) 014004.
   \\
   Y.~Kanazawa and Y.~Koike,
   Phys.\ Lett.\ B {\bf 478} (2000) 121;
   Phys.\ Lett.\ B {\bf 490} (2000) 99. 


\bibitem{model-h1}
   P.~V.~Pobylitsa and M.~V.~Polyakov,
   Phys.\ Lett.\ B {\bf 389} (1996) 350 [arXiv:hep-ph/9608434].\\
   P.~Schweitzer, D.~Urbano, M.~V.~Polyakov, C.~Weiss, P.~V.~Pobylitsa 
   and K.~Goeke,
   Phys.\ Rev.\ D {\bf 64} (2001) 034013 [arXiv:hep-ph/0101300].

\bibitem{Boer:2003cm}
   D.~Boer, P.~J.~Mulders and F.~Pijlman,
   arXiv:hep-ph/0303034.

\bibitem{Koike:2002ti}
   Y.~Koike,
   Talk given at 15th International Spin Physics Symposium (SPIN 2002), 
   Long Island, New York, 9-14 Sep 2002, 
   arXiv:hep-ph/0210396. 

\bibitem{Anselmino:1999pw}
  M.~Anselmino, M.~Boglione and F.~Murgia,
  Phys.\ Rev.\ D {\bf 60} (1999) 054027
  [arXiv:hep-ph/9901442].

\bibitem{Metz:iz}
   A.~Metz,
   Phys.\ Lett.\ B {\bf 549} (2002) 139. 

\bibitem{Brodsky:2002pr}
   S.~J.~Brodsky, D.~S.~Hwang and I.~Schmidt,
   Phys.\ Lett.\ B {\bf 553} (2003) 223;
   Nucl.\ Phys.\ B {\bf 642} (2002) 344.
   D.~Boer, S.~J.~Brodsky and D.~S.~Hwang,
   Phys.\ Rev.\ D {\bf 67} (2003) 054003.

\bibitem{Gamberg:2003ey}
   L.~P.~Gamberg, G.~R.~Goldstein and K.~A.~Oganessyan,
   Phys.\ Rev.\ D {\bf 67} (2003) 071504
   [arXiv:hep-ph/0301018]. 

\bibitem{Henneman:2001ev}
   A.~A.~Henneman, D.~Boer and P.~J.~Mulders,
   Nucl.\ Phys.\ B {\bf 620} (2002) 331 [arXiv:hep-ph/0104271].

\bibitem{Efremov:2002ut}
   A.~V.~Efremov, K.~Goeke and P.~Schweitzer,
   Phys.\ Rev.\ D {\bf 67} (2003) 114014
   [arXiv:hep-ph/0208124]. 

\bibitem{Jackson:1989ph}
  J.~D.~Jackson, G.~G.~Ross and R.~G.~Roberts,
  Phys.\ Lett.\ B {\bf 226} (1989) 159.

\bibitem{Efremov-in-prep}
   A.~V.~Efremov, K.~Goeke and P.~Schweitzer, {\sl in preparation}.

\bibitem{Diakonov:1983hh}
   D.~Diakonov and V.~Y.~Petrov,
   Nucl.\ Phys.\ B {\bf 245} (1984) 259.

\bibitem{Diakonov:1987ty}
   D.~Diakonov, V.~Y.~Petrov and P.~V.~Pobylitsa,
   Nucl.\ Phys.\ B {\bf 306} (1988) 809.

\bibitem{Christov:1995vm}
   For a review see: C.~V.~Christov {\it et al.},
   Prog.\ Part.\ Nucl.\ Phys.\  {\bf 37} (1996) 91 [arXiv:hep-ph/9604441].

\bibitem{Dressler:1999hc}
   B.~Dressler and M.~V.~Polyakov,
   Phys.\ Rev.\ D {\bf 61} (2000) 097501 [arXiv:hep-ph/9912376].

\bibitem{Pobylitsa:2002fr}
   P.~V.~Pobylitsa, 
   arXiv:hep-ph/0212027.



\bibitem{Diakonov:1996sr}
   D.~Diakonov, V.~Petrov, P.~V.~Pobylitsa, M.~V.~Polyakov and C.~Weiss,
   Nucl.\ Phys.\ B {\bf 480} (1996) 341 [arXiv:hep-ph/9606314].

\bibitem{Diakonov:1995qy}
   D.~Diakonov, M.~V.~Polyakov and C.~Weiss,
   Nucl.\ Phys.\ B {\bf 461} (1996) 539 [arXiv:hep-ph/9510232].

\bibitem{Collins:2003fm}
   J.~C.~Collins,
   Acta Phys.\ Polon.\ B {\bf 34} (2003) 3103 [arXiv:hep-ph/0304122].


\end{thebibliography}
\end{document}